\documentclass[prl,twocolumn]{revtex4}

     \usepackage{bm}
     \usepackage{graphicx}
     \usepackage{amssymb}
     \usepackage{amsmath}
     \usepackage{eufrak}
     \usepackage{color}
     \usepackage[utf8]{inputenc} 
     \usepackage{hyperref}
     \usepackage{pifont}
     \usepackage{ulem}
     \usepackage{verbatim}

\allowdisplaybreaks
			
\newcommand{\nix}[1]{}
\begin{document}

\title{MIRO-like oscillations of magneto-resistivity\\ in GaAs heterostructures induced by THz radiation}

\author{
T.~Herrmann,$^1$
I.~A. Dmitriev,$^{2,3}$ D.~A.~Kozlov,$^{4,5}$  M.~Schneider,$^{1}$ B.~Jentzsch,$^1$
 Z.~D.~Kvon,$^{4,5}$ P.~Olbrich,$^1$ V.~V.~Bel'kov,$^3$ A. Bayer,$^1$ D. Schuh,$^1$
D.~Bougeard,$^1$  T. Kuczmik,$^1$ M. Oltscher,$^1$ D.~Weiss,$^1$
and S.~D.~Ganichev$^{1}$
}

\affiliation{$^1$ Terahertz Center, University of Regensburg, 93040 Regensburg, Germany}
\affiliation{$^2$ Max Planck Institute, 70569  Stuttgart, Germany}
\affiliation{$^3$ Ioffe Institute, 194021 St.\,Petersburg, Russia}
\affiliation{$^4$ Rzhanov Institute of Semiconductor Physics, 630090 Novosibirsk, Russia }
\affiliation{$^5$ Novosibirsk State University, 630090 Novosibirsk, Russia }

\begin{abstract}	
We report on the study of terahertz radiation induced MIRO-like oscillations of
magneto-resistivity in GaAs heterostructures. Our experiments provide an answer on two most
{  intriguing} questions -- effect of radiation helicity and the role of the edges --
yielding crucial information for understanding of the MIRO origin. Moreover, we demonstrate
that the range of materials exhibiting radiation-induced magneto-oscillations can
be largely extended by using high-frequency radiation.
\end{abstract}

\maketitle{}

\section{Introduction}

One of the most stunning phenomena discovered in the past decade in
two dimensional electron systems (2DES) is microwave (MW) induced
magneto-resistance oscillations (MIRO)\,\cite{Zudov01,Ye01,Mani02,Zudov03,Yang03,Dorozhkin03,Smet05}, reviewed, e.g., in \cite{Dmitriev12}. These oscillations are, as Shubnikov-de Haas oscillations (SdH), periodic on a $1/B$ scale, but occur at lower magnetic fields and show much weaker temperature dependence. Phenomenologically, they are very similar to
Weiss oscillations\,\cite{Weiss1989}, which reflect the commensurability between the cyclotron orbit radius and the period of a periodic potential. MIRO by contrast, reflect the commensurability between the MW photon energy $2\pi \hbar f$ and the cyclotron energy $\hbar \omega_c$. Here, $\hbar$ is the reduced Planck constant, $f$ the MW and $\omega_c$ the cyclotron frequencies. In extremely clean samples the minima of the MIRO develop into zero resistance
states \cite{Mani02,Zudov03,Yang03}  which are explained \cite{Andreev03} in terms of an instability of the system and formation of current domains, occurring when the conductivity  becomes negative under MW irradiation (see also \cite{Dmitriev12,Dmitriev13,Dorozhkin15}).

\begin{figure}[h]
\includegraphics[width=0.8\linewidth]{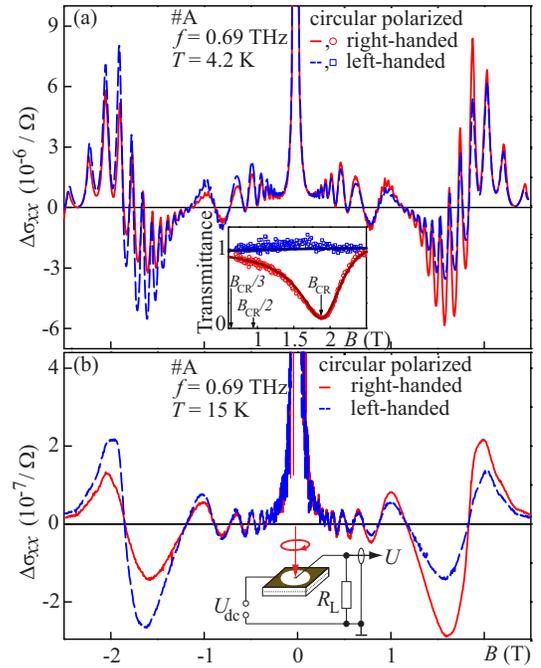}
\caption{Magnetic field dependence of
$\Delta\sigma_{xx}$  induced by modulated \textit{cw}
radiation with $f=0.69$~THz in sample~\#A.
The dependencies are obtained for right-
and left-handed
circularly polarized radiation.
The inset in panel (a) presents the transmission data.
Solid lines show the transmission calculated after\,\cite{Abstreiter} by
taking multiple reflections within the substrate
and the superradiant decay into account\,\cite{Abstreiter,Chiu76,Mikhailov04,Kono2014}.
}
\label{Fig1}
\end{figure}

In spite of numerous experiments and significant
advances in their theoretical understanding,
there is still no commonly accepted microscopic description of the effect\,\cite{Zudov15,Mikhailov15} and
the ongoing MIRO investigations remain challenging\,\cite{Shi13,Levin14,Chakraborty14,Shi15,Levin15,Bykov16}.
Consequently new materials have been
studied\,\cite{Zudov14,Shi14,Yamashiro15,Karcher16}
and new theoretical
models have been put forward \,\cite{Beltukov16,Volkov14,Zhirov13,Mikhailov11,Chepelianskii09}. To the most pressing issues which need to be clarified experimentally and which might help to differentiate between the different models belong the MIRO polarization dependence and the
``bulk'' or ``edge''  nature of the effect.

So far the majority of experimental work in this direction has been done in the MW regime (1-350 GHz) and there are only a few
reports on MIRO excited at terahertz (THz) frequencies\,\cite{Wirthmann2007,Kvon2013,Mani2013}.
Here we report on the observation of pronounced MIRO-like oscillations
induced by THz radiation, i.e. in the frequency regime between 0.7 and 2\,THz.
We exploit the specific advantages of THz laser radiation
not present in the MW regime, i.e., the
possibility to focus
it
onto a spot
smaller than the sample's size and easy control of the radiation's polarization  state.
The most important features clearly detected on a large variety of samples are (i) a very weak dependence of the oscillations' amplitude  on the photon helicity and (ii) the
``bulk'' nature of the effect.
Furthermore, our study shows that the  MIRO
oscillations can be excited at THz frequencies even in the samples with
low mobility whereas in the MW  range ultra-high mobility samples are crucially needed
for this type of experiments.

We study the radiation induced oscillations in
AlGaAs/GaAs quantum wells\,(QW)
of $10$ or $12.5$\,nm thickness.  While THz MIRO-like oscillations
have been observed in Hall bar samples\,\cite{Kvon2013}
in this work we used Corbino disk
samples,
which measure directly $\sigma_{xx}$.
The inner, $r_i$, and outer radius, $r_o$, of the gold-germanium contacts
and samples' transport parameters are given in Table\,\ref{sample},
 see also Suppl. Materials.
The effect of THz radiation on conductivity $\sigma_{xx}$ was studied in
the temperature range from $T =2.5$\,K to about 20\,K. The MIRO-like effect is most pronounced after the cooled samples have been exposed to room light, but the effect is also present for samples kept in the dark. All data shown here were taken after illuminating the samples.

The THz experiments were performed using  continuous wave ($cw$)
CH$_2$O$_2$ and
CH$_2$F$_2$ lasers\,\cite{3aa,Olbrich2009,DMS2}
operating at frequencies $f = 0.69$
and $1.63$\,THz, respectively.
The power, $P$, being of the order of 10\,mW at the sample surface,
has been controlled by  pyroelectric detectors.
The radiation at normal incidence is focused
onto a spot size of about 3.4\,mm diameter. The beam
 has an almost Gaussian profile, measured by
a pyroelectric camera\,\cite{Ganichev1999,Ziemann2000}.
Right- and left-handed circularly polarized radiation is obtained by a $\lambda/4$
quartz plate\,\cite{book,BelkovSSTlateral}.
The inset in Fig.\,\ref{Fig1}(b) sketches the 
set-up.
Applying a
bias voltage, $U_{\rm dc}=50$\,mV,
and measuring the voltage 
$U$  on the load resistor $R_L = 50$\,Ohm we obtained the radiation induced change of the sample conductivity $\Delta\sigma_{xx} \propto U$.

\begin{table}
\centering
\begin{tabular}{|c|c|c|c|c|c|c|}
	\hline
	  Sample          &$r_i$ &$r_o$ &$\mu$              &$n_e$                &$\tau_p$&$\tau_q$\\
                    &  mm  &   mm &10$^3$\,[cm$^2/$Vs]&10$^{11}$\,[cm$^{-2}$]&ps      &ps \\
\hline
		\#A
							& 0.25 & 1.5  & 820               & 12.0                 & 33     & 4.2\\
	  \#B
								& 0.3  & 1.0  & 1800              & 9.3                  & 71     & 12.0\\
	
	  \#C
								& 0.3  & 1.0  & 150               & 18.0                 & 6      & 1.3\\
		\#D
							& 0.25 & 1.5  & 980               & 24.0                 & 39     & 1.9\\
	
		 \#E
							& 0.25 & 1.5  & 280               & 3.7                  & 11     & 4.1\\
		\hline
\end{tabular}
\caption{Samples and their transport data at $T=2$\,K  including the electron density $n_e$ and mobility $\mu$, as well as the momentum, $\tau_p$, and quantum, $\tau_q$, relaxation times.}
\label{sample}
\end{table}


\begin{figure}
\includegraphics[width=0.9\linewidth]{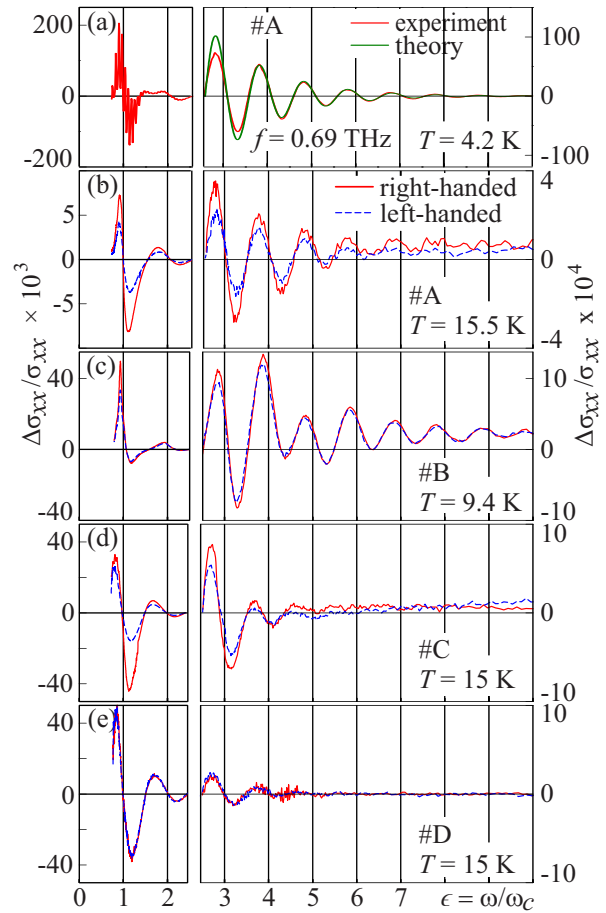}
\caption{  THz radiation induced conductivity change
$\Delta\sigma_{xx}/\sigma_{xx}$ as a
function of $\omega / \omega_c$.
The green line in (a) is a fit of the low-$B$ tail of the oscillations using Eq.\,(\ref{MIRO}) with a $B$-independent $A_\epsilon=A_\infty$
 \cite{footnote_phase}.
Such fit yields $A_\infty^{exp}=0.056$ in full agreement with theory\,\cite{Dmitriev05} yielding $A_\infty^{th}\sim 0.04$.
The monotonic part of $\Delta\sigma_{xx}/\sigma_{xx}$ in panel (a) is subtracted.}
\label{Fig2}
\end{figure}

Irradiating the samples with THz radiation and sweeping the magnetic field $B$ results in conductivity
oscillations $\Delta \sigma_{xx}$ starting already at  $|B| \sim 0.1$\,T. Figure\,\ref{Fig1} shows the data for sample \#A and $f = 0.69$\,THz.
The oscillations reflect two $1/B$ periodicities: one at low $B$ corresponding to the MIRO-like oscillations and
the other, at higher fields $|B|\sim 1$\,T, corresponding to SdH oscillations.
While the former oscillations are detected up to temperatures of about 20\,K the latter {  are} observed at substantially lower temperatures only.
The oscillations are  observed for a large number of structures with mobilities between
$\mu=1.8\times10^6$
and $1.5 \times 10^5$\,cm$^2/$Vs and with  electron densities from
 $n_e=4$ to $24 \times 10^{11}$\,cm$^{-2}$. Figure\,\ref{Fig2} shows $\Delta\sigma_{xx}$  normalized {  to} the dark conductivity, $\Delta\sigma_{xx}/\sigma_{xx}$, for different samples. The data clearly show that the oscillations are periodic in $\epsilon = \omega /\omega_c$ with $\omega=2\pi f$ and
exhibit extrema at $\epsilon = \epsilon_N \pm1/4$, where $\epsilon_N\simeq N=1,\,2,\,3\ldots$ denote the position of the nodes~\cite{footnote_phase} and $+(-)$ corresponds to minima (maxima).
Here, the position of the cyclotron resonance (CR) at $\omega = \omega_c$  has been obtained from the transmission experiments displayed in the inset of Fig.\,\ref{Fig1}\,(a).
The overall behavior of the observed $1/B$--oscillations including the exponential damping at low $B$ and
the $1/4$-shift of the minima and maxima from the nodes
corresponds to the well known MIRO effects as is further discussed below.
Strikingly, Figs.\,\ref{Fig1} and\,\ref{Fig2} reveal that
the  oscillation amplitudes depend only weakly on the radiation helicity and  some
samples
(\#B
and \#D)
%
show nearly no helicity dependence.
By contrast, transmission, displayed in the inset in
Fig.\,\ref{Fig1} (a), strongly depends on helicity. CR for positive $B$ occurs only for right-handed circular polarization and vice versa.
Indeed, in agreement with the theory of CR\,\cite{bookCR}, the transmitted power drops
nearly to zero for the cyclotron resonance active configuration
(CRA, red data in the inset),
whereas
 the signal for the opposite helicity (blue data), i.e. for the inactive configuration (CRI),  is almost $B$-independent.
Active and inactive helicity interchange for negative magnetic fields (not shown).
Note that the detected polarization behavior and the shape of transmittance minimum are well fitted by taking  multiple
reflections
in the
substrate and superradiant decay into account\,\cite{Abstreiter,Chiu76,Mikhailov04,Kono2014},
see solid lines in the inset of Fig.\,\ref{Fig1} (a) and Suppl.
Materials.
To  demonstrate that the weak dependence of the MIRO-like oscillations
on the radiation helicity is not caused by  saturation effects we
measured $\Delta \sigma_{xx}$ at different levels of radiation intensity.
These data demonstrate that the signals depend linearly on the radiation intensity
and reducing the intensity by an order of magnitude  does not change the
ratio between  CRA and CRI signals for all values of $\omega/\omega_c$.
The overall same results are obtained for the higher frequency $f = 1.63$\,THz.
The only  differences are the detected signal magnitude and rescaled
$B-$field positions of CR and oscillations.

\begin{figure}
\includegraphics[width=0.9\linewidth]{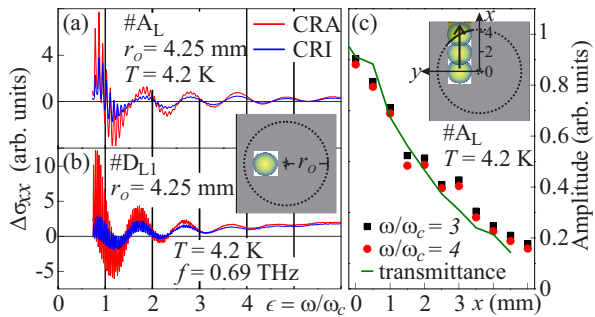}
\caption{ (a) and (b)
$\Delta\sigma_{xx}$ as a function of   $\omega / \omega_c$ for large size ($r_o = 4.25$\,mm) samples
\#A$_{\rm L}$
and \#D$_{\rm L1}$.
A metal mask (inset, grey square) with a hole on top of the Corbino device prevents illumination of the contacts.
(c) Oscillation amplitude as a function of the laser spot position using a metal mask with a slit (see inset).
The beam is scanned along the slit from ``bulk''  regions of the Corbino disk onto the outer contact area.
The solid line shows the transmission signal
measured simultaneously with the oscillations.
}
\label{Fig3}
\end{figure}

\begin{figure}
\includegraphics[width=0.9\linewidth]{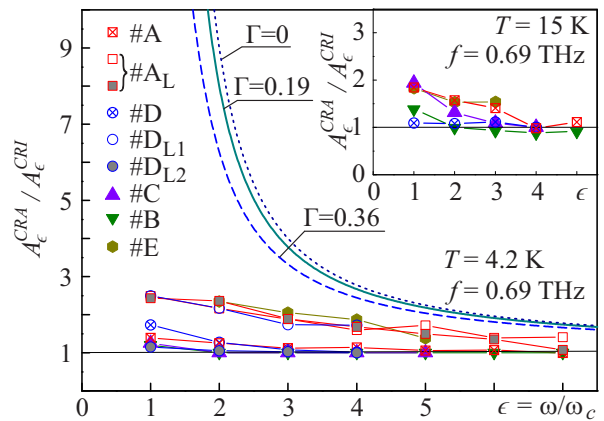}
\caption{ $A_\epsilon^\text{CRA}/A_\epsilon^\text{CRI}$ vs. $\epsilon$ for all samples.
For sample \#A$_{\rm L}$ it
is shown for the laser focused in between the contacts
(open squares) and onto the edge of the outer Corbino contact (solid squares).
\#D$_{\rm L1}$ and \#D$_{\rm L2}$ (open and solid circles) are samples of same size, made from the same wafer.
Dashed, green, and dotted lines are calculated using
{  Eq.\,(\ref{ratio})} for different
 values of $\Gamma \propto n_e$. $\Gamma = 0.19 $ and $\Gamma = 0.36 $ correspond to parameters taken from samples
 \#A and \#D, respectively.
}
\label{Fig4}
\end{figure}

The results shown in Figs.\,\ref{Fig1} and \ref{Fig2} are obtained for Corbino discs with
an outer diameter 3\,mm, thus being of the order of the laser spot size.
To exclude  possible effects stemming from illumination of the contacts'
edges we carried out measurements on discs with 8.5~mm diameter,
i.e. significantly larger than the spot size. These samples are indicated by a subscript ${\rm L}$ in the sample index.
Figure\,\ref{Fig3} (a) and (b) show that the oscillations can be efficiently excited without illuminating the contacts.
We measured   the photoconductivity (PC)   while illuminating the sample through a square-shaped
hole in a metal mask covering the contacts. The metal film was mounted
at a distance of about 1\,mm from the sample surface in a way that only the
2DES was illuminated, see inset in Fig.\,\ref{Fig3}(a).
To prove that the boundaries of the Corbino electrodes do not contribute
to the PC
signal we removed a part of the metal screen and
{  scanned} the beam across the edge, see inset in Fig. \,\ref{Fig3}(c).
Figure \,\ref{Fig3}(c) shows that the signal
decreases
as the beam { approaches}
the edge, demonstrating that
the oscillation amplitude just follows the decrease in absorbed radiation power, probed via the transmission
simultaneously with { the PC} signal.
The data clearly show that the edges do not contribute to the  oscillations
and, consequently, ponderomotive forces excited by edge { illumination\,\cite{Mikhailov11} } as well as
radiation-induced modification of the edge { transport\,\cite{Chepelianskii09} } are not essential for MIRO generation, at least in the THz regime.
Importantly, the  polarization dependence of the oscillations remains weak
for any position of the laser spot {  including
both geometries, where the contacts are avoided and
where
the beam is focused on the edge.}
This is {  illustrated} in Fig.\,\ref{Fig4}, where the ratios of the oscillation amplitude
for CRA and CRI configurations are shown as a function of  $\omega / \omega_c$.
In all cases and especially for low oscillation indices the experimental CRA/CRI ratio
is always much smaller than the calculated ones, shown for different radiative decay
$\Gamma$ (discussed below) in Fig.\,\ref{Fig4}.
The finding that the THz induced oscillations depend only weakly on helicity is in line with results of
 Smet et al., who examined MIROs at MW frequencies\,\cite{Smet05}.

Now we discuss the results within the most commonly used theoretical approaches to explain MIRO,
i.e., the inelastic\,\cite{Dorozhkin03,Dmitriev03,Dmitriev05} and the displacement\,\cite{Ryzhii70,Ryzhii86,Durst03,Vavilov04,Khodas08} mechanisms,
which  ascribe the magneto-oscillations of the PC to radiation-assisted scattering between
disorder-broadened Landau levels. This theory is a nonequilibrium extension
of linear transport theory  in high Landau levels\,\cite{Ando}, and accounts
for both, real-space displacements of electron orbits in individual photon-assisted
scattering events (displacement contribution) and associated nonequilibrium
occupation of electronic states within the disorder-broadened Landau levels
(inelastic contribution). Within this theory the conductivity oscillation are given by \cite{Dmitriev09}
\begin{equation}\label{MIRO}
\Delta \sigma_{xx}/\sigma_{xx} = -\epsilon A_\epsilon \sin(2\pi \epsilon)\exp(-\alpha \epsilon),
\end{equation}
where
the exponential damping with $\alpha=2\pi/\omega\tau_q$ corresponds
to the dirty limit $\omega_c\tau_q\ll 1$ of strongly overlapping Landau levels.
Importantly, the factor $A_\epsilon$ becomes $B$-independent for $\epsilon=\omega/\omega_c \gg 1$.
Fitting the low-$B$ data in Fig.\,\ref{Fig2}(a)
using Eq.\,(\ref{MIRO}) with constant $A_\infty=A_\epsilon|_{\epsilon\to\infty}$ reveals that the experimental oscillation amplitude,
$A_\infty^{exp}=0.056$, agree well with the theoretical predictions for the
inelastic mechanism\,\cite{Dmitriev05},  $A_\infty^{th}\sim 0.04$,
see\,\cite{footnote_phase} and Suppl. Materials for details.
This shows, in particular, that  $\Delta\sigma_{xx}$ follows
 the $n_e^3/\omega^{4}$ dependence predicted for a {  dominant} inelastic mechanism.
We emphasize that all parameters required for this comparison are extracted from the independent measurements of transmission and
magneto-transport.
In accordance with Eq.\,(\ref{MIRO}), the number of resolved oscillations for all samples in Fig.\,\ref{Fig2}
is roughly equal to $\omega\tau_q/2$.  Here $\tau_q$ are taken from the analysis of SdH oscillations, see Table\,\ref{sample}.
This highlights the important role of $\tau_q$ in the appearance of oscillations and explains
why at THz frequencies they can be observed in samples with relatively low mobility. Thus, the THz regime opens perspectives
for the observation of MIRO in broader range of materials.

While {  the inelastic mechanism of MIRO describes  the
experimental findings for large $\omega/\omega_c$ quite well
it can not explain the weak polarization dependence measured for  $\omega/\omega_c  \leq 5$.}
In theory, the polarization dependence of  $\Delta\sigma_{xx}$
follows the absorbance $K =\sigma^{(\pm)}_{xx}(\omega)E_\pm^2/2$, where $\sigma^{(\pm)}_{xx}(\omega)$ is the dissipative part of the linear dynamic conductivity,  $E_\pm$ is the amplitude of the screened radiation field acting on the electrons, and $+$($-$) labels the CRI (CRA) circular polarization. Although
the magneto-oscillations in $\Delta\sigma_{xx}$
are of quantum origin,   their polarization dependence
stems from  Drude theory and reflects the classical dynamics
of 2DES in crossed magnetic field $B$ and electric field of the THz wave, yielding \cite{bookCR}
\begin{equation}
\label{CR}
A_\epsilon \propto K \propto \dfrac{E_\pm^2}{(1\pm|\omega_c|/\omega)^2+\gamma^2}, \quad \gamma= 1/\omega\tau_p\ll 1.
\end{equation}
In order to relate $E_\pm$, acting on the electrons in 2DES, to the electric field $E^{(0)}_\pm$ of the incoming wave, one needs to take into account both the dynamical screening by the 2DES
and multiple reflections within the dielectric substrate (Fabry-Perot interference). The former is dominated by the Hall part $\sigma^{(\pm)}_{xy}(\omega)$ of the dynamic conductivity
and is described by the superradiant decay rate $\Gamma=e^2 n_e/2\omega\epsilon_0 m c\gg\gamma$\,\cite{Abstreiter,Chiu76,Mikhailov04,Kono2014}.
The effect of the dynamical screening is most simply expressed in the case of constructive Fabry-Perot interference \cite{footnote_interference}:
the absorbance still has the form of Eq.~(\ref{CR}) where $E_\pm$ get replaced by $E^{(0)}_\pm$ and $\gamma$ by $(\Gamma+\gamma)$.
Then, since $\Gamma\gg \gamma$, the
ratio of  CRA and CRI absorption coefficient
and oscillation amplitude is given by
\begin{equation}
\dfrac{K^\text{CRA}}{K^\text{CRI}}  = \dfrac{A_\epsilon^\text{CRA}}{A_\epsilon^\text{CRI}}\simeq\dfrac{(1+|\omega_c|/\omega)^2+\Gamma^2}{(1-|\omega_c|/\omega)^2+\Gamma^2}.
\label{ratio}
\end{equation}
Note that the value of $\Gamma$ is fully determined by $n_e$ and $m$ known precisely from independent measurements. Taking multiple reflections into account (see Suppl. Material), the transmission data for CRA and CRI polarization can be perfectly fitted using the above expression for $\Gamma$, see inset in Fig.\,\ref{Fig1}(a). In Figure\,\ref{Fig4} we present the ratio (\ref{ratio}) calculated for different samples.
The smallest $A_\epsilon^\text{CRA}/A_\epsilon^\text{CRI}$ value is obtained for the largest $\Gamma$, in our case $\Gamma = 0.36$ for sample
\#D with the largest carrier density $n_e=2.4\times10^{12}\ \text{cm}^{-2}$, see
dashed line in Fig.\,\ref{Fig4}. Clearly, for {low harmonics  $\omega/\omega_c \leq 5$, the theoretical predictions} for $A_\epsilon^\text{CRA}/A_\epsilon^\text{CRI}$ strongly deviate from the experimental data in Fig.\,\ref{Fig4}.
Unreasonably high
values of $\Gamma \geq 1.5$ would be needed to match experiment.
For higher harmonics, however, i.e., for $\omega/\omega_c \gtrsim 5$,  $A_\epsilon^\text{CRA}/A_\epsilon^\text{CRI}\rightarrow 1$  both in theory and experiment. As detailed in the Suppl. Material,
the data in this low-$B$ region are in good quantitative agreement with theory.

To summarize, our experiments provide clear evidence that MIRO-like oscillations in the THz regime are excited in the ``bulk''  of 2DES and
not at contacts/sample boundaries. In the THz range they are
also observed at much lower mobilities of the 2DES as compared
to the MW regime suggesting that they appear at $\omega\gtrsim \tau_q^{-1}$.
Although at high harmonics of the cyclotron resonance the observed MIRO-like oscillations are well described by the inelastic mechanism \cite{Dmitriev05}, their polarization dependence
at low harmonics is at odds with any existing theoretical description of MIRO.

\acknowledgments  Financial  support via the SFB 689 of the German Science Foundation DFG, RFBR (N 14-02-01246) and RAS are gratefully acknowledged.



\begin{thebibliography}{99}
\bibitem{footnote_phase} In our experiments, the actual positions of the nodes $\epsilon_N=-\epsilon_0+N$ deviate
from the integer values $\epsilon_N=N=1,\,2,\,3,\ldots$ established in both theory and previous MW measurements \cite{Dmitriev12}.
The small shifts $\epsilon_0$ vary from sample to sample both in sign and magnitude, see Suppl. Materials, and should be disregarded.
In the example shown in Fig.~\ref{Fig2}a, the best fit is obtained with $\epsilon_0=-0.09$.
\bibitem{footnote_interference} Equation (\ref{ratio}) is written for the constructive Fabry-Perot interference inside the dielectric substrate, in which case the broadening of the CR is maximized. In other conditions the ratio $A_\epsilon^\text{CRA}/A_\epsilon^\text{CRI}$ can only increase, see Suppl. Material.
\end{thebibliography}
\end{document}


\title{Supplementary material { for ``MIRO-like} oscillations of magneto-resistivity in GaAs heterostructures induced by THz radiation''}

\author{
T.~Herrmann,$^1$
I.~A. Dmitriev,$^{2,3}$ D.~A.~Kozlov,$^{4,5}$  M.~Schneider,$^{1}$ B.~Jentzsch,$^1$
 Z.~D.~Kvon,$^{4,5}$ P.~Olbrich,$^1$ V.~V.~Bel'kov,$^3$ A. Bayer,$^1$ D. Schuh,$^1$
D.~Bougeard,$^1$  T. Kuczmik,$^1$ M. Oltscher,$^1$ D.~Weiss,$^1$
and S.~D.~Ganichev$^{1*}$
}

\affiliation{$^1$ Terahertz Center, University of Regensburg, 93040 Regensburg, Germany}
\affiliation{$^2$ Max Planck Institute, 70569  Stuttgart, Germany}
\affiliation{$^3$ Ioffe Institute, 194021 St.\,Petersburg, Russia}
\affiliation{$^4$ Rzhanov Institute of Semiconductor Physics, 630090 Novosibirsk, Russia }
\affiliation{$^5$ Novosibirsk State University, 630090 Novosibirsk, Russia }



\maketitle{}

\section{Linear magneto-transport}

All samples were characterized using linear magnetotransport measurements both before and after illumination by the room light. Standard lock-in technique has been applied with the excitation ac current of 1-100~nA and frequency of 12~Hz. The sheet density $n_s$ and the quantum lifetime $\tau_q$ in Table~I of the main text were extracted from data obtained in the two-point Corbino geometry. The mobility $\mu$ from Corbino measurements is not reliable due to possible significant contribution of the contact resistance in the vicinity of zero magnetic field. For this reason we additionally conducted standard four-point Van der Pauw measurements for each GaAs wafer under investigation and extracted the values of $\mu$ and $\tau_p$ (listed in Table~I of the main text) from the values of sheet resistance $\rho_{xx}$.

\begin{figure}
\includegraphics[width=0.9\linewidth]{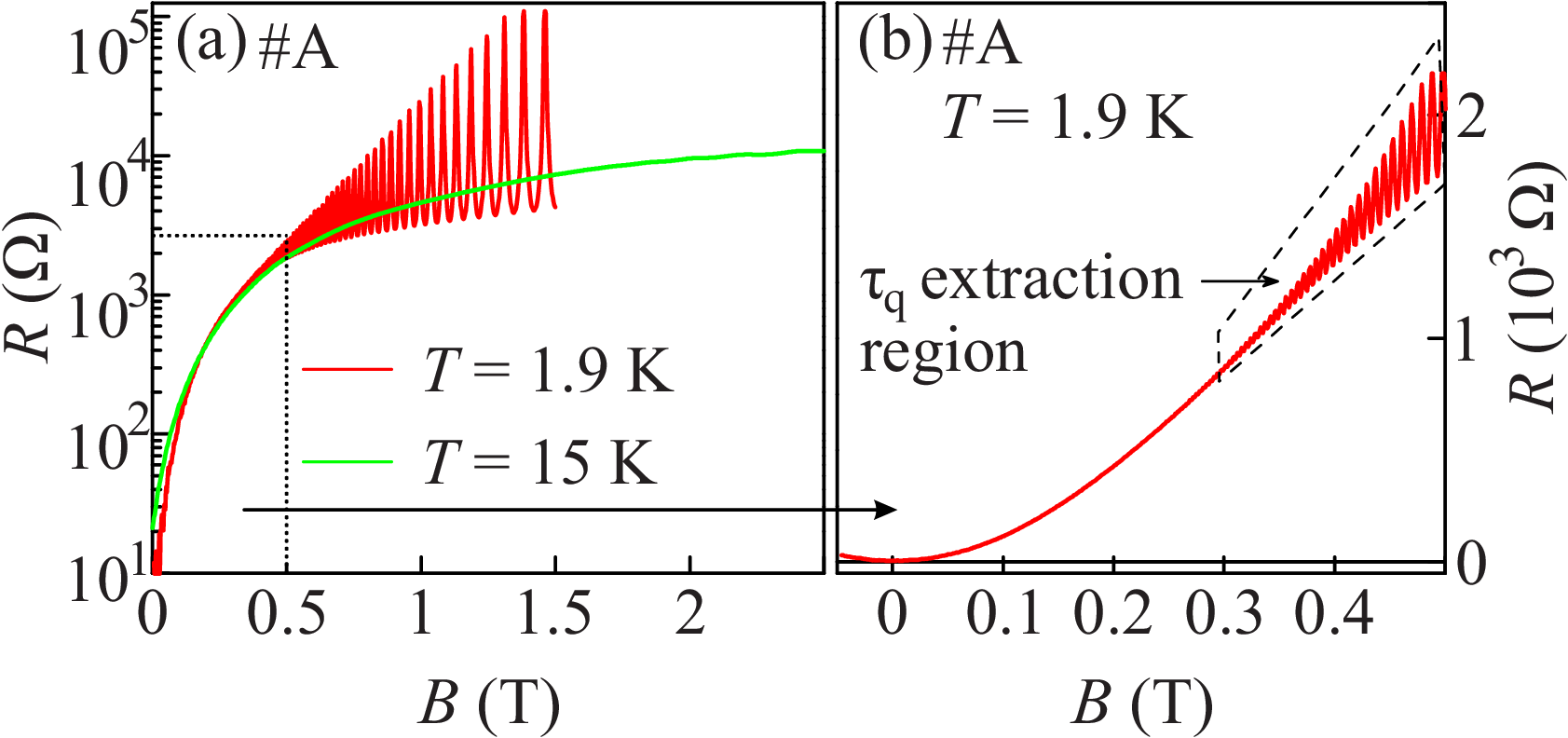}
\caption{Panel (a): Corbino magnetoresistance $R(B)$ measured on sample 506B at the temperatures $T$=1.9\,K and 15\,K. Panel (b): the same data (for $T$=1.9K) in the range of small $B$ where the SdH oscillations are not very pronounced. The dashed line marks the region of $B$ used for fitting with the Lifshits-Kosevich formula which yields the value of $\tau_q$ in Table~I of the main text.}
\label{FigSup1}
\end{figure}

An example of measured dependence of the magnetoresistance (inverse conductance) $R(B)$ in the Corbino geometry for two values of temperature $T$ is shown in the Fig.~\ref{FigSup1}(a). At $T=15$~K one observes only classical parabolic magnetoresistance. The curve measured at $T=1.9$~K shows similar behavior at low magnetic fields with the SdH oscillations superimposed at $B>0.3$~T. The electron sheet density was obtained from the periodicity of SdH oscillations in the reciprocal magnetic field assuming spin-degenerated Landau levels. The values of the quantum life time $\tau_q$ were extracted using the Lifshits-Kosevich (LK) formula from the fitting of the low-field SdH oscillations measured at low $T$, see Fig.~\ref{FigSup1}(b). For each $R(B)$ trace we calculated the monotonous part $\langle R(B)\rangle$ by smoothing the original curve. Next, we calculated the relative oscillatory signal $\delta r(B) = [R(B)-\langle R(B)\rangle]/\langle R(B)\rangle$ which was fitted using the LK formula in the range $|\delta r(B)| < 0.3$ where the LK formula is applicable [$B < 0.5$~T in Fig.~\ref{FigSup1}(b)].

\section{Determination of the terahertz field acting on 2D electrons using the transmission data}

A precise control of the amplitude and polarization of the high frequency field acting on electrons in 2DES is extremely challenging in the microwave range of frequencies used in MIRO experiments and, so far, was only realized using a unique quasi-optical set-up in Ref.~\cite{Smet05}. By contrast, in the terahertz range one has an opportunity to implement well-defined focused laser beams of incoming radiation. After a simple analysis of the supplementary transmission data {  similar to that presented in Refs.~\cite{Abstreiter,Chiu76,Mikhailov04,Kono2014},} one can determine the screened terahertz field $E_\pm$ acting on electrons in the bulk of the 2DES. This knowledge, in turn, enables (i) a quantitative comparison of the experimental results for MIRO-like oscillations with theoretical predictions and (ii) a systematic comparison of the data obtained from different samples with varying density and transport parameters.

\begin{figure}
\includegraphics[width=0.9\linewidth]{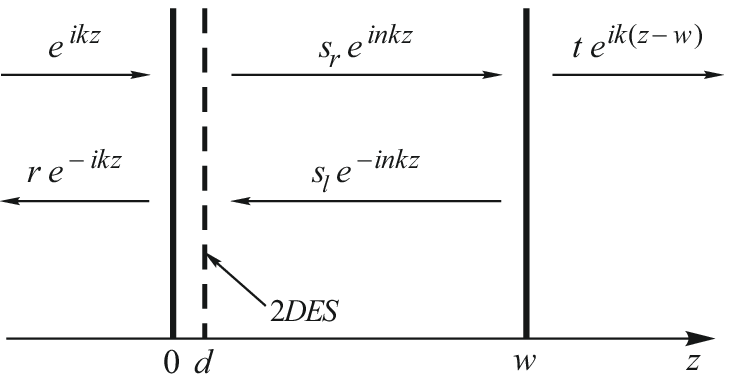}
\caption{Illustration of the model used to relate the dynamic screening and transmission. Horizontal arrows indicate the partial reflected/transmitted THz waves, vertical solid lines at $z=0$ and $z=w$ -- the front and back interfaces of the sample (dielectric slab with the refractive index $n=3.6$) with vacuum, dashed vertical line at $z=d \ll 1/nk$ -- the position of the 2DES characterized by the dynamic conductivity $\hat{\sigma}(\omega)$. The amplitudes $r$ and $t$ are the reflectance and transmittance amplitudes, while the amplitude $s=s_r+s_l=1+r$ describes the reduction of screened terahertz field in the plane of 2DES with respect to the  incident field.}
\label{FigSup2}
\end{figure}

To relate the screening and transmission, we consider a circularly polarized wave normally incident from $z<0$ upon a sample which occupies $0<z<w$ and contains a 2DES at $z=d$, see Fig.~\ref{FigSup2}. In our samples the 2DES is located close to the interface at $z=0$, such that $d\ll 1/n k$, which allows us to set $d=0$. Here $k=\omega/c$ is the wave vector in vacuum and $n=3.6$ is the refraction index of the GaAs substrate.  By contrast, the optical path $\phi=n k w$ across the whole sample including the substrate is large, $\phi\gg 1$, which necessitates the account for multiple reflections between the dielectric interfaces (Fabry-Perot interference).

The electric field of the circularly-polarized incident wave is given by
\begin{equation}
\textbf{E}_{i}=E_{i}\text{Re}\ \textbf{e}_\pm e^{i k z-i\omega t},
\end{equation}
where the complex vectors $\sqrt{2}\textbf{e}_\pm=\textbf{e}_x \pm i \textbf{e}_y$ describe the helicity of the wave. The structure is modeled as a dielectric slab with the refraction coefficient $n\simeq 3.6$ at $0<z<w$ placed in vacuum and a conducting plane at $z=d$ characterized by the dynamic conductivity $\hat{\sigma}(\omega)$. Assuming isotropic transport in the 2DES, $\sigma^{(\pm)}_{xx}(\omega)=\sigma^{(\pm)}_{yy}(\omega)$ and $\sigma^{(\pm)}_{yx}(\omega)=-\sigma^{(\pm)}_{xy}(\omega)$, one finds that the dynamic conductivity tensor $\hat\sigma(\omega)$ is diagonal in the chiral basis,
\begin{equation}\label{chiral}
\hat\sigma(\omega) {\bf e}_\pm=\sigma_\pm {\bf e}_\pm,\quad  \sigma_\pm=\sigma^{(\pm)}_{xx}(\omega)\pm i\sigma^{(\pm)}_{xy}(\omega).
\end{equation}
When the time-reversal symmetry is broken, $\sigma_{xy}\neq 0$, the polarization state of the incoming and transmitted/reflected waves is in general different. Still, the axial symmetry for the normally incident circularly polarized wave and isotropic 2DES is preserved, therefore the individual circular components $\propto {\bf e}_+$ and ${\bf e}_-$ do not mix and can be considered independently. We write down the electric field in the whole space as $\textbf{E}(z,t)=E_{i}\text{Re}\ \textbf{e}_\pm a(z) e^{-i\omega t}$, with $a(z)=e^{i k z}+ r e^{-i k z}$ at $z<0$, $a(z)=s_{r} e^{i k z}+ s_{l} e^{-i k z}$ at $0<z<w$, and $a(z)=t e^{-i k (z-w)}$ at $z>w$, see Fig.~\ref{FigSup2}. Here $t$ and $r$ are the transmittance and reflectance amplitudes, while $s=s_{r}+ s_{l}$ gives the amplitude $E_\pm$ of the screened field $\text{Re}\ \textbf{e}_\pm E_\pm e^{-i\omega t}=E_{i}\text{Re}\ \textbf{e}_\pm s e^{-i\omega t}$ acting on electrons in the 2DES. Indices $\pm$ of the amplitudes $a,~t,~s,~r$ are suppressed for better readability.

Introducing the vector $A(z)=(a, \partial_z a)^T$ and implementing the boundary conditions at the interfaces provided by the Maxwell equations, one obtains
\begin{equation}\label{maxwell}
A(\delta)=M_e A(-\delta),\quad A(w+\delta)=M_\text{d}A(\delta).
\end{equation}
Here $\delta$ is a positive infinitesimal,
\begin{equation}\label{AA}
A(-\delta)=\binom{1+r}{i k (1-r)},\quad A(w+\delta)=\binom{t}{i k t},
\end{equation}
\begin{equation}\label{AA}
A(\delta)=\binom{s_{r}+ s_{l}}{i n k (s_{r}-s_{l})}.
\end{equation}
The transfer matrix
\begin{equation}
\label{M_e}
M_e=
\begin{pmatrix}
  1     & 0\\
 2 i \omega \sigma_\pm/\epsilon_0 c^2 & 1
\end{pmatrix}
\end{equation}
describes the discontinuity of $\partial_z a$ associated with the ac current induced in the 2DES, while
\begin{equation}
\label{M_e}
M_d=
\begin{pmatrix}
  \cos\phi     & (n k)^{-1}\sin\phi\\
 -n k\sin\phi & \cos\phi
\end{pmatrix},\qquad \phi=n k w,
\end{equation}
describes propagation of the wave inside the dielectric slab.

Solution to the above equations reads
\begin{equation}\label{t}
t^{-1}=(1+\tilde{\sigma}_\pm)\cos\phi-i\dfrac{1+n^2+2\tilde{\sigma}_\pm}{2n}\sin\phi,
\end{equation}
where $\tilde{\sigma}_\pm=\sigma_\pm/2\epsilon_0 c$ (in Gaussian units, $\tilde{\sigma}_\pm=2\pi\sigma_\pm/c$).
The screened field in the 2DES plane is given by
\begin{equation}\label{s}
s=1+r=t(\cos\phi-in^{-1}\sin\phi).
\end{equation}
Note that the helicity dependence in $t,~r$ and $s$ is solely due to $\sigma_\pm$ entering Eqs.~(\ref{t}) and (\ref{chiral}).

As discussed below, for our purposes here it is sufficient to describe the ac conduction in the 2DES within the classical Drude approximation. The Drude expression for the ac conductivity,
\begin{equation}\label{drude}
\sigma_\pm=\dfrac{e^2 n_e/m }{\tau_p^{-1}-i(\omega\pm\omega_c)},
\end{equation}
immediately follows from the Drude equation for the drift velocity
\begin{equation}\label{newton}
(\partial_t+1/\tau_p)\mathbf{v}=-e/m(\mathbf{E}+\mathbf{v}\times\mathbf{B}).
 \end{equation}
This equation describes the classical dynamics of an electron subjected to crossed ac electric field $\textbf{E}$ and constant magnetic field $\textbf{B}$
in the presence of a damping force $-m \mathbf{v}/\tau_p$ associated with scattering by impurities. Using Eq.~(\ref{drude}), we represent the dimensionless conductivity $\tilde{\sigma}$ entering Eq.~(\ref{t}) in the form
\begin{equation}\label{tilde}
\tilde{\sigma}_\pm=\dfrac{\sigma_\pm}{2\epsilon_0 c}=\dfrac{\Gamma}{\gamma-i(1\pm\omega_c/\omega)}.
\end{equation}
Here the dimensionless parameter
\begin{equation}
\label{Gamma}
\Gamma=\dfrac{e^2 n_e}{2 \epsilon_0 c m\omega}=\dfrac{0.301\ \text{T}}{B_\text{cr}}\ \dfrac{n_e}{10^{12}\ \text{cm}^{-2}}
\end{equation}
represents the so called radiative or superradiant decay. It is fully determined by the electron density $n_e$ and the position of the cyclotron resonance $B_{cr}=\omega m/e$. The other parameter
\begin{equation}
\label{gamma}
\gamma=\dfrac{1}{\omega\tau_p}=\dfrac{1}{B_\text{cr}[\text{T}]\mu[10^4\ \text{cm}^{2}/\text{V s}]}
\end{equation}
represents the momentum relaxation in the 2DES. The classical Drude formula (\ref{drude}) underestimates the width of the cyclotron resonance in $\sigma^{(\pm)}_{xx}(\omega)=\text{Re}~\sigma_\pm$ in the quantum regime of well separated Landau levels. However, in conditions of our experiments (i) $\omega\tau_q\gtrsim 1$, meaning the quantum corrections are not yet significant, (ii) the screening and transmission are dominated by $\sigma_{xy}(\omega)$ which is insensitive to the Landau quantization as $\gamma\ll \Gamma\lesssim 1$.

Substitution of Eq.~(\ref{tilde}) into Eq.~(\ref{t}) yields the power transmission coefficient {  $T=|t|^2$~\cite{Abstreiter},}
\begin{widetext}
\begin{equation}\label{Tfull}
T_\pm=\dfrac{y^2 + \gamma^2}{[y^2+(\gamma + \Gamma)^2]\cos^2\phi+[\tilde n^2 y^2+(\tilde n\gamma
+ \Gamma/n)^2]\sin^2\phi-(\tilde n-n^{-1})y\Gamma\sin2\phi},\quad y=1\pm \dfrac{\omega_c}{\omega},\;\tilde{n}=\dfrac{n+n^{-1}}{2}.
\end{equation}
\end{widetext}
With known parameters $\Gamma$ and $\gamma$ determined from the electron density $n_e$ and mobility $\mu$, fitting of the $B$-dependence of transmission data using Eq.~(\ref{Tfull}) provides the interference parameter $\phi=n k w$ (up to an irrelevant integer part of $\phi/\pi$) and the effective mass entering $\omega_c/\omega=B/B_{cr}$. For $\gamma\ll \Gamma\lesssim 1$, the position of the cyclotron resonance $B_{cr}=m\omega/e$ coincides with the minimum in transmission for the CRA polarization. In the example shown in Fig.~1 of the main text for the sample \#A ($n_e=1.2\times10^{12}\text{ cm}^{-2}$ and $\mu=82\times10^4\text{ cm}^{2}/\text{V\,s}$ corresponding to $\Gamma=0.19$ and $\gamma=0.0065$) this procedure gives $\phi/\pi=-0.04$ and $B_{cr}=1.88$~T. Note a precise agreement for both CRA and CRI polarizations obtained in a single scan including both positive and negative $B$.

Using the relation (\ref{s}), with known $T_\pm$ and the interference parameter $\phi$ one readily determines the intensity $|E_\pm|^2$ of the screened terahertz field acting on 2D electrons relative to the intensity of the incoming wave,
\begin{equation}
\label{ss}
|E_\pm|^2/E_i^2=|s_\pm|^2=T_\pm (\cos^2\phi +n^{-2}\sin^2\phi).
\end{equation}
This knowledge, in turn, enables a quantitative comparison of experimental data with theoretical predictions. To illustrate the importance of the interference effects, we address the range of magnetic field $|\omega_c|\ll \omega$ where $y$ entering Eq.~(\ref{Tfull}) is close to unity irrespective of the polarization of the wave.  For relevant $\gamma,~\Gamma\ll 1$, in this limit the constructive interference ($\cos^2\phi=1$) produces $T_\pm=|E_\pm|^2/E_i^2=1$, while the destructive interference ($\sin^2\phi=1$) yields a reduced value of $T_\pm=4 n^2/(n^2+1)^2\simeq 0.27$ and much stronger reduction of $|E_\pm|^2/E_i^2=4 /(n^2+1)^2\simeq 0.02$. This shows that without knowledge of $\phi$ no quantitative comparison of experiment with theory, as well as comparison of experimental results obtained from different samples can be made as long as the magnitude of the MIRO effect is concerned.


\section{Comparison of experimental results for MIRO-like oscillations with theoretical predictions}

Within theory combining the displacement and inelastic mechanisms of MIRO, the relative change of the diagonal dc conductivity induced by circularly polarized radiation is given {  by~\cite{Dmitriev09}}
\begin{equation}\label{theory}
\dfrac{\Delta\sigma_{xx}}{\sigma_{xx}}=-\eta P S_\pm \dfrac{2\pi\omega}{\omega_c}\sin\dfrac{2\pi\omega}{\omega_c}\exp\left(-\dfrac{2\pi}{\omega_c\tau_\text{q}}\right).
\end{equation}
Here the dimensionless radiation power $P$ and the polarization-dependent factor $S_\pm$ are
\begin{equation}\label{PS}
P=\dfrac{2\pi e^2 n_e E_i^2}{m^2\omega^4},\quad S_\pm=\dfrac{|s_\pm|^2}{\gamma^2+ (1\pm\omega_c/\omega)^2},
\end{equation}
[with the screening factor $|s_\pm|^2$ given by Eq.~(\ref{ss})]. The quantum scattering time $\tau_q$ determines the width of Landau levels, and the factor $\eta=\eta_\text{in}+\eta_\text{dis}$ contains the mechanism-specific parameters of the 2DES. Namely, the inelastic contribution,
\begin{equation}\label{etain}
\eta_{in}=\dfrac{2\tau_\text{in}}{\tau_p}\simeq\dfrac{2\pi\hbar^3 n_e}{m\tau_p T^2},
\end{equation}
includes the rate of inelastic electron-electron collisions $\tau_\text{in}^{-1}\propto T^2$, while the temperature-independent displacement contribution $\eta_\text{dis}$ is fully determined by the elastic scattering. The displacement contribution  is highly sensitive to the correlation properties of disorder and obtains the maximum value $\eta_\text{dis}=3/2$ when the momentum relaxation is dominated by large angle scattering from short-range impurities. The exponential damping of oscillations in Eq.~(\ref{theory}) corresponds to the regime of overlapping Landau levels relevant to our current experiments. Three amplitude factors in Eq.~(\ref{theory}) combine into $A_\epsilon=2\pi\eta P S_\pm $ entering Eq.~(1) in the main text.

\textit{Helicity dependence.} One immediately observes that the polarization dependence in Eq.~(\ref{theory}) is the same as in the classical Drude absorption coefficient, given by $K=\sigma^{(\pm)}_{xx} (\omega)|E_\pm|^2/2$, see Eqs.~(\ref{drude}) and (\ref{ss}). The reason is that magnetooscillations in both mechanisms originate from the photon-assisted transitions between disorder-broadened Landau levels. While being quantum in nature, they include matrix elements of optical transitions reflecting the classical dynamics in crossed fields described by Eq.~(\ref{newton}). According to Eqs.~(\ref{Tfull}), (\ref{ss}), and (\ref{PS}), the minimum ratio of the MIRO-like signal for CRA and CRI polarizations is expected in the case of constructive interference, when
\begin{equation}\label{Sc}
S_\pm=\dfrac{1}{(\Gamma+\gamma)^2+ (1\pm\omega_c/\omega)^2}, \quad \cos^2\phi=1.
\end{equation}
Assuming for definiteness $\omega_c>0$ and taking into account that $\gamma\ll\Gamma$, we obtain for this ratio the expression quoted in the main text:
\begin{equation}\label{AA}
\dfrac{A_\epsilon^\text{CRA}}{A_\epsilon^\text{CRI}}=\dfrac{K^\text{CRA}}{K^\text{CRI}}=\dfrac{(1+\omega_c/\omega)^2+\Gamma^2}{(1-\omega_c/\omega)^2+\Gamma^2}.
\end{equation}

Figure 4 in the main text demonstrates a striking disagreement between the experimental data showing weak helicity dependence of MIRO-like oscillations and theory (\ref{AA}) predicting much stronger oscillations near the lowest harmonics $\omega/\omega_c=1,~2,~3$ for the active circular polarization. Even with the largest $\Gamma\simeq 0.36$ realized in the sample \#D with the highest electron density $n_e=2.4\times10^{12}$~cm$^{-2}$, this ratio at $\omega/\omega_c\sim 1,\;2,\;3$ is by far larger than that observed in the experiment. In samples with lower density or different values of $\phi$, the theoretical ratio $A_\epsilon^\text{CRA}/A_\epsilon^\text{CRI}$ can only increase. For instance, in the opposite case of destructive interference the width $\Gamma+\gamma$ entering Eq.~(\ref{Sc}) is replaced with considerably smaller value $\varkappa\Gamma+\gamma$, where $\varkappa=2/(n^2+1)\simeq 0.14$,
\begin{equation}\label{Sd}
S_\pm=\dfrac{\varkappa^2}{(\varkappa\Gamma+\gamma)^2+ (1\pm\omega_c/\omega)^2}, \quad \sin^2\phi=1.
\end{equation}
Near higher harmonics, at $\omega\gg|\omega_c|$, oscillations become polarization-independent in both theory and experiment. With the interference parameter $\phi$ extracted from the transmission data, for the corresponding low-$B$ tail of $\Delta\sigma_{xx}$ one can perform a quantitative comparison of theory and experiment. Such analysis, presented below, shows that other theoretical predictions including the magnitude of oscillations are in good agreement with our data, which makes the polarization puzzle even more dramatic.

\textit{Magnitude of oscillations.}
In theory, the amplitude $A_\epsilon=2\pi\eta P S_\pm$ becomes $B$-independent (and, therefore, polarization-independent) at $\epsilon=\omega/\omega_c\gg 1$, where it can be replaced by $A_\infty=A_\epsilon|_{\epsilon\to\infty}$.
As illustrated in Fig.~2(a) { of the main text}, the data collected for the sample \#A at $T=4.2$~K for $\omega/\omega_c>3$ can be perfectly fitted using the expression
\begin{equation}
\label{fit}
\dfrac{\Delta\sigma_{xx}}{\sigma_{xx}}=-{ A_\infty^{exp} }\epsilon\sin(2\pi[\epsilon+\epsilon_0])\exp(-\alpha\epsilon),\
\end{equation}
where $\epsilon=\omega/\omega_c=B_{cr}/B$ and a phenomenological constant $\epsilon_0$ introduced to account for an additional phase shift absent in theory.
This fit provides $B_{cr}=1.84$~T consistent with the value $B_{cr}=1.88$~T extracted from the transmission data, the phase shift $\epsilon_0=-0.09$, the damping parameter $\alpha=2\pi/\omega\tau_q=0.97$, and the amplitude of oscillations $A_\infty^{exp}=0.056$. Note that similar analysis for other samples yields a small phase shift $\epsilon_0$ of varying sign and magnitude, see Table~\ref{sup}. Therefore, we believe that finite $|\epsilon_0|\ll 1$ has an instrumental origin and should not be regarded as conflicting with the well established position of the nodes, $\epsilon_0=0$, reported earlier for the microwave range of frequencies~\cite{Dmitriev12}.

\begin{figure}
\includegraphics[width=0.9\linewidth]{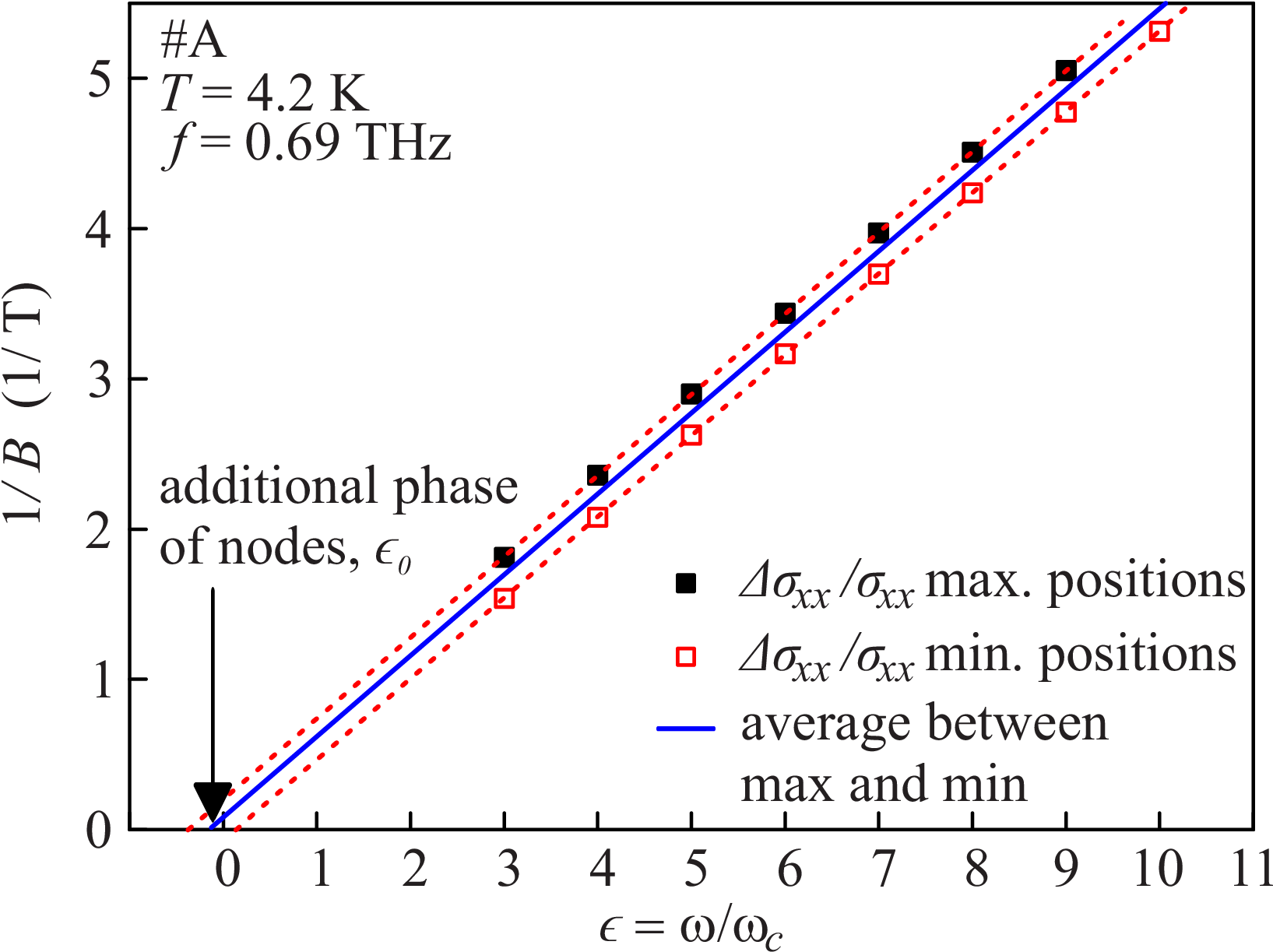}
\caption{The inverted values of the $B$-positions corresponding to the minima and maxima of oscillations for sample \#A at $T=4.2$~K near different harmonics of the CR (integer $\omega/\omega_c=N=3,~4,~5,\cdots$) plotted against $\omega/\omega_c=B_{cr}/B$. The data points fall on straight lines with the slope $B_{cr}=1.84$~T. The median line, marking the positions of the nodes, crosses the horizontal axis at $\epsilon_0=-0.09$ instead of $\epsilon_0=0$ expected from theory and previous MIRO measurements. At the same time, the horizonal distance between the lines through the minima and maxima is 1/2 which reproduces the 1/4 phase shift of the minima and maxima reported before. The fitting procedure involving four parameters $B_{cr}$, $\epsilon_0$, $\alpha$, and $A_\infty$ is nevertheless reliable since many oscillation periods are resolved, see also Fig.~\ref{FigSup4}.
}
\label{FigSup3}
\end{figure}

\begin{figure}
\includegraphics[width=0.9\linewidth]{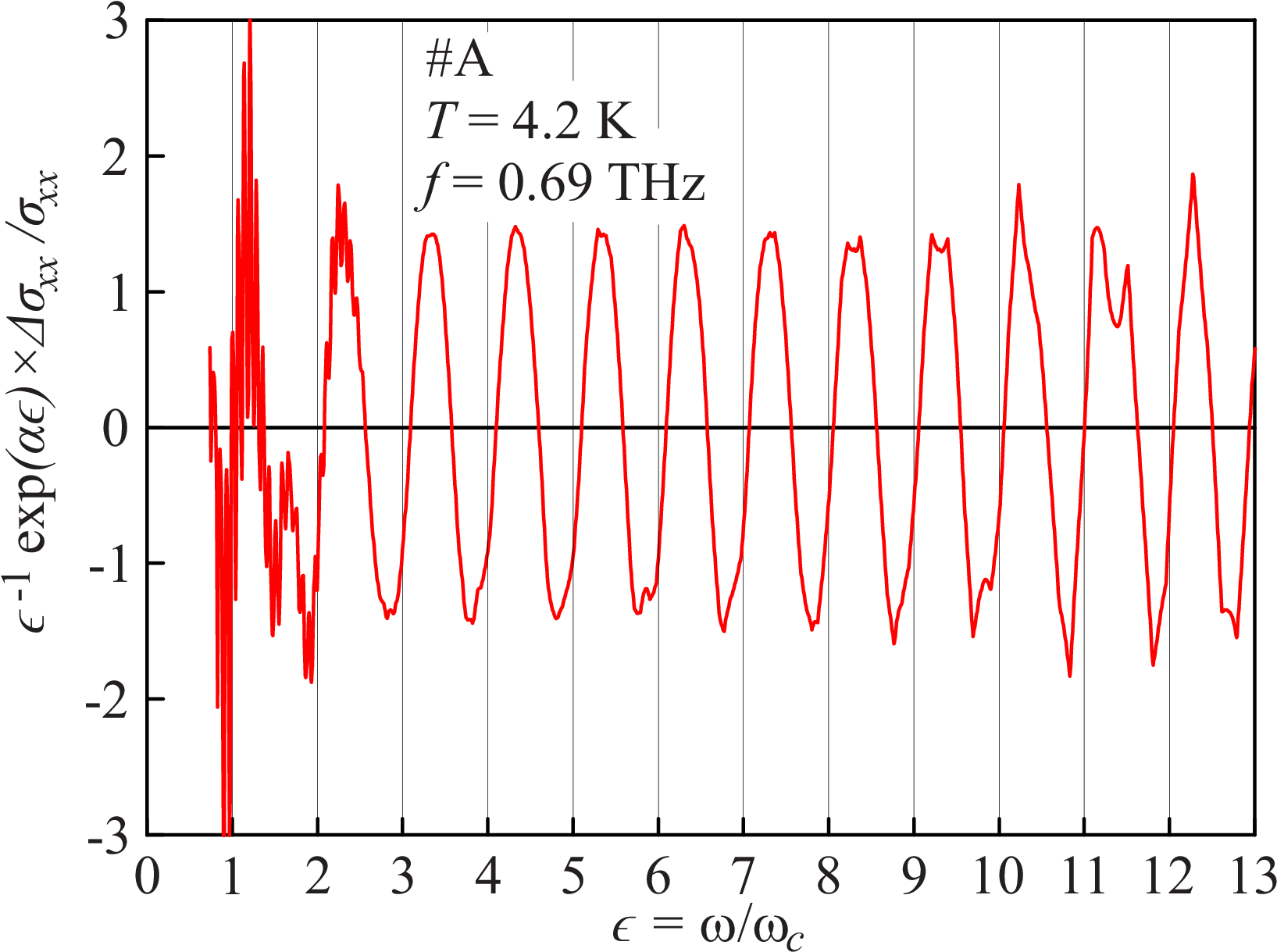}
\caption{The $\omega/\omega_c$-dependence of the relative magnetooscillations $\Delta\sigma_{xx}/\sigma_{xx}$ with removed smooth background when multiplied by $\epsilon^{-1}\exp(\alpha\epsilon)$ with $\alpha=0.97$ obtained from the fit to Eq.~(\ref{fit}). This plot illustrates the accuracy of the fit and quality of the data at large harmonics of the CR ($\epsilon=\omega/\omega_c\gg 1$). In samples where smaller number of oscillations is resolved, the accuracy of the fit is reduced. It is still sufficient to demonstrate an overall agreement of the magnitude of oscillations with theoretical predictions, see Table~\ref{sup}. }
\label{FigSup4}
\end{figure}

Figures~\ref{FigSup3} and  \ref{FigSup4} illustrate that the fitting procedure is reliable in spite of several fitting parameters.
In Fig.~\ref{FigSup3}, we extract the $B$-positions of the minima and maxima of oscillations near different harmonics of the CR (integer $\omega/\omega_c=N=3,~4,~5,\cdots$); the inverted values of extracted $B$-positions are plotted against $N$. They fall on straight lines with the slope $B_{cr}=1.84$~T thus confirming strict $1/B$-periodicity of oscillations. The median crosses the $\epsilon$-axis at $\epsilon_0=-0.09$ while both theory and previous MIRO measurements yield $\epsilon_0=0$, corresponding to $\Delta\sigma_{xx}\propto -\sin(2\pi B_{cr}/B)$. The value of the additional phase $\epsilon_0=-0.09$ is not consistent with measurements on other samples and should be disregarded. Note that the relative phase shift between minima and maxima (the horizontal distance between the solid lines in Fig.~\ref{FigSup3}) is exactly 1/2 as expected in the regime of exponentially damped oscillations. The accuracy of the exponential damping at low $B$ is further illustrated in Fig.~\ref{FigSup4}, where oscillations are multiplied by $\epsilon^{-1}\exp(\alpha\epsilon)$ with $\alpha=0.97$ extracted from the fitting. We mention that the value of $\tau_q$ corresponding to $\alpha=0.97$ is about 3 times shorter than the value extracted from the SdH oscillations in the same sample at lower temperature, see also Ref.~\cite{Studenikin07}. In the microwave range, by contrast, $\tau_q$ extracted from the low-$B$ damping of MIRO is usually found to be larger than $\tau_q$ obtained from the SdH oscillations. It was attributed to long-range potential fluctuations which lead to an additional damping of the SdH oscillations but do not modify the MIRO amplitude~\cite{Dmitriev12}.

On the other hand, the amplitude $A_\infty$ can be estimated using Eq.~(\ref{theory}), which gives
\begin{equation}
\label{A}
A_\infty^{th}=2\pi\eta P S_\infty\simeq \dfrac{(2\pi\hbar)^3 e^2 n_e^2 E_i^2}{m^3\tau_p T^2\omega^4}\simeq 0.04
\end{equation}
Here we took into account that for $\phi=-0.04$ extracted from the transmission data for \#A (very close to the constructive interference condition, $\cos^2\phi=0.985$) and for $\Gamma=0.19$ the polarization factor $S_\infty=S_\pm|_{\omega\gg\omega_c}$ is close to unity. In other words, the screened terahertz field $E_\pm$ away from the cyclotron resonance here is almost identical to the incident field $E_i$ of the wave. The electric field in the center of the 10~mW Gaussian beam with the half-width $r_0=1.7$~mm at half-maximum is $7.6$~V/cm, which gives the average intensity  across the sample area $\langle E_i^2\rangle= 45~\text{V}^2/\text{cm}^2$. With $T=4.2$~K and $m=0.074~m_0$ corresponding to $B_{cr}=1.84$~T extracted from the period of oscillations, Eq.~(\ref{etain}) gives $\eta_\text{in}\simeq 12$, therefore the displacement contribution $\eta_\text{dis}\leq 3/2$ can be neglected. Given the accuracy of the above estimate, the resulting amplitude of oscillations,  $A_\infty^{th}\sim 0.04$, coincides with the value $A_\infty^{exp}=0.056$ extracted from the experimental data.

Similar fitting using Eq.~(\ref{fit}) and theoretical estimates using Eq.~(\ref{theory}) combined with the analysis of the transmission data were performed for other samples. The results are presented in Table~\ref{sup}.  A precise determination of the dependence on the electron mobility, density, and temperature, and on radiation power and frequency requires collection of additional data and further refinement of the measurement techniques, and is relegated to future work.

\begin{table}
\centering
\begin{tabular}{|r|r|r|r|r|r|r|r|}
	\hline
	Sample & $T$\,[K] & $B_{cr}$\,[T]& $\epsilon_0$ & $\alpha$ & $10^3\cdot A_\infty^{exp}$ & $10^3\cdot A_\infty^{th}$ & $\phi/\pi$ \\
	\hline
	\hline
	\#A 
& 4.2 & 1.8 & -0.09 & 1 & 56 & 40 & -0.04 \\
	\#B 
& 9.4 & 1.8 & -0.09 & 0.6 & 1.3 & 1.7 & -0.1\\
	\#C 
& 15 & 1.8 & 0.09 & 2.6 & - & 5 & 0.15\\
	\#D 
& 15 & 1.9 & 0.01 & 1.6 & 24 & 8 & -0.07 \\
	\#E 
& 4.2 & 1.8 & -0.05 & 1.2 & 20 & 9 & -0.1 \\
	\hline
\end{tabular}
\caption{Parameters of the samples ($B_{cr}$, $\epsilon_0$, $\alpha$, and $A_\infty^{exp}$) extracted from fitting of oscillations measured on different samples to Eq.~(\ref{fit}). The theoretical value $A_\infty^{th}$ is calculated using Eq.~(\ref{theory}), with the interference parameter $\phi$ extracted from the transmission data. Due to small number of resolved oscillations,
the parameter $A_\infty^{exp}$ could not
be reliably extracted for the sample \#C. }
\label{sup}
\end{table}